\documentstyle[12pt]{article}
\begin{document}
\tolerance=5000
\def\pp{{\, \mid \hskip -1.5mm =}}
\def\cL{{\cal L}}
\def\be{\begin{equation}}
\def\ee{\end{equation}}
\def\bea{\begin{eqnarray}}
\def\eea{\end{eqnarray}}
\def\tr{{\rm tr}\, }
\def\nn{\nonumber \\}
\def\e{{\rm e}}
\def\D{{D \hskip -3mm /\,}}

\def\SEH{S_{\rm EH}}
\def\SGH{S_{\rm GH}}
\def\AdS5{{{\rm AdS}_5}}
\def\S4{{{\rm S}_4}}
\def\gfv{{g_{(5)}}}
\def\gfr{{g_{(4)}}}
\def\SC{{S_{\rm C}}}
\def\RH{{R_{\rm H}}}

\  \hfill 
\begin{minipage}{3.5cm}
NDA-FP-78 \\
July 2000 \\
\end{minipage}

\vfill

\begin{center}
{\large\bf 
Can We Live on the Brane in Schwarzschild-anti de Sitter Black Hole?}

\vfill

{\sc Shin'ichi NOJIRI}\footnote{email: nojiri@cc.nda.ac.jp} 
and {\sc Sergei D. ODINTSOV}$^{\spadesuit}$\footnote{
email: odintsov@ifug5.ugto.mx, odintsov@mail.tomsknet.ru},

\vfill

{\sl Department of Applied Physics \\
National Defence Academy, 
Hashirimizu Yokosuka 239, JAPAN}

\vfill

{\sl $\spadesuit$ 
Tomsk State Pedagogical University \\
634041 Tomsk, RUSSIA \\
and \\
Instituto de Fisica de la Universidad de 
Guanajuato \\
Apdo.Postal E-143, 37150 Leon, Gto., MEXICO} 

\vfill

{\bf ABSTRACT}

\end{center}
The model of d5 higher derivative (HD) gravity admitting Schwarzschild-
-anti de Sitter Black Hole (S-AdS BH) as exact solution is considered.
The surface counterterms are added to the complete action, they are fixed 
by the condition of finiteness of bulk AdS spacetime when brane goes to
infinity. As a result the brane (observable Universe) is defined dynamically 
 in terms of d5 theory parameters (brane tension is fixed). Brane radius is
always bigger than horizon radius and the 4d Universe itself could be
static or inflationary. Natural ganeralization of this model gives warped
compactification (with dynamically generated brane) in the next-to-leading
order of AdS/CFT correspondence.

\newpage

Some years ago, in the remarkable paper \cite{RU} 
(see also related idea in ref.\cite{Akama}) the following question
has been addressed: Do we live on the domain wall in the 
multidimensional Universe? The title of the paper \cite{RU} almost 
literally coincides with this question. Recently, we understood 
that coupling of multidimensional gravity with brane where observer 
lives may occur in such a way that gravity on the brane is 
trapped \cite{RS}. As a result the explosion 
of the interest in brane-world cosmology \cite{CH,cosm}, 
brane-world black holes \cite{SNOO} and its possible 
manifestations \cite{exp} took place.
Moreover, it has been realized that there are deep connections 
between  brane-world scenarios and AdS/CFT correspondence \cite{AdS} 
in string theory (see books \cite{GSW,Pol} for a review) where warped
compactification naturally introduces brane-world set-up.
In such situation it is really interesting 
from various points of view to address the questions:
Do we live on the brane? and What is the bulk? Clearly this goes 
as the development of idea of ref.\cite{RU}.

 In the present Letter 
 in frames of the model of higher derivative (HD) gravity we 
are trying to construct d5 S-AdS BH where four-dimensional brane 
(observable Universe) is realized dynamically. 
d4 brane Universe could be the static one or inflationary Universe.
Brane radius defined dynamically is always bigger than horizon radius 
what indicates that observer lives outside the horizon. Before the
explanation of
details of our model the important remark is in order. In the most standard
brane-world scenarios \cite{RS,CH,cosm} bulk (Einstein) gravity contains
bulk cosmological constant as free parameter and couples with the brane.
Adding Gibbons-Hawking boundary term \cite{3} which makes the variational 
procedure to be well-defined and brane cosmological constant term (brane 
tension) one is left with second free parameter, i.e. brane tension.
Playing with these two parameters one can construct various brane-world
configurations already in Einstein gravity. 

Our approach lies on the different line \cite{sergio} which is more
suitable in terms of 
AdS/CFT correspondence. Namely, one considers the addition of surface
counterterms (instead of addition of free tension brane) to complete action.
Then, brane tension is not free parameter anymore (as well as other surface
counterterms parameters), they are fixed by the condition of finiteness of 
bulk spacetime when brane goes to infinity. Clearly, in such approach 
the brane-world scenarios are much more restrictive. Nevertheless, as we
will see the theory parameters like coefficients of HD terms may help in
the construction of 
interesting brane-worlds where brane is defined dynamically. That is the
reason why we consider HD gravity where one has additional parameters of
theory (the coefficients of HD terms in 5d action). Moreover, as we explain
at the end such  d5 S-AdS BH in HD gravity has natural extension as   
warped compactification in next-to-leading order of AdS/CFT correspondence.

The general action of $d+1$-dimensional $R^2$ gravity is given by:
\be
\label{vi}
S=\int d^{d+1} x \sqrt{-\hat G}\left\{a \hat R^2 
+ b \hat R_{\mu\nu}\hat R^{\mu\nu}
+ c \hat R_{\mu\nu\xi\sigma}\hat R^{\mu\nu\xi\sigma}
+ {1 \over \kappa^2} \hat R - \Lambda \right\}\ .
\ee
It is well-known that such theory is multiplicatively renormalizable 
\cite{ste} and asymptotically free (see book \cite{BOS} for a review) in
four dimensions.
The following conventions of curvatures are used
\bea
\label{curv}
R&=&g^{\mu\nu}R_{\mu\nu} \nn
R_{\mu\nu}&=& -\Gamma^\lambda_{\mu\lambda,\kappa}
+ \Gamma^\lambda_{\mu\kappa,\lambda}
- \Gamma^\eta_{\mu\lambda}\Gamma^\lambda_{\kappa\eta}
+ \Gamma^\eta_{\mu\kappa}\Gamma^\lambda_{\lambda\eta} \nn
\Gamma^\eta_{\mu\lambda}&=&{1 \over 2}g^{\eta\nu}\left(
g_{\mu\nu,\lambda} + g_{\lambda\nu,\mu} - g_{\mu\lambda,\nu} 
\right)\ .
\eea
When $a=b=c=0$, the action (\ref{vi}) becomes that of the 
Einstein gravity. 
When $c=0$ \footnote{For non-zero $c$ such S-AdS BH solution may be
constructed perturbatively\cite{SNO}. It is useful to establish the  
higher-derivative AdS/CFT correspondence\cite{SNO} and find the 
strong coupling limit of super Yang-Mills theory with two 
supersymmetries in next-to-leading order.}, 
Schwarzschild-anti de Sitter space is an exact solution:
\bea
\label{SAdS}
ds^2&=&-\e^{2\rho_0}dt^2 + \e^{-2\rho_0}dr^2 
+ r^2\sum_{i,j}^{d-1} g_{ij}dx^i dx^j\ ,\nn
\e^{2\rho_0}&=&{1 \over r^{d-2}}\left(-\mu + {kr^{d-2} \over d-2} 
+ {r^d \over l^2}\right)\ .
\eea
Here $\mu$ is the parameter corresponding to mass and the scale 
parameter $l$ is given by solving the following equation:
\be
\label{ll}
0={d^2(d+1)(d-3) a \over l^4} + {d^2(d-3) b \over l^4} \nn
- {d(d-1) \over \kappa^2 l^2}-\Lambda\ .
\ee
We also assume $g_{ij}$ expresses the Einstein manifold, 
defined by $r_{ij}=kg_{ij}$, where $r_{ij}$ is the Ricci tensor 
defined by $g_{ij}$ and $k$ is the constant. 
For example, if $k>0$ the boundary can be 4 dimensional 
de Sitter space (sphere when Wick-rotated), if $k<0$, anti-de Sitter 
space or hyperboloid, or if $k=0$, flat space. 
In the following, we Wick-rotate the time coordinate by replacing 
$t\rightarrow it$. Then in order to avoid the conical singularity 
in the Wick-rotated spacetime, the time coordinate 
$t$ should have a period ${1 \over T}$, which is given 
for $d=4$ by
\be
\label{T}
T={\sqrt{{k^2 \over 4} + {4\mu \over L^2}} 
\over 2\pi \sqrt{-{k \over 2} + \sqrt{{k^2 \over 4} 
+ {4\mu \over L^2}}}}\ .
\ee
Here $T$ is the Hawking temperature.

One suppose that there is a boundary at $r=r_0$, where brane 
lies. Then we need to add boundary term, especially 4 dimensional 
cosmological term in order to get brane-world Universe, 
i.e. RS scenario for 4d gravity \cite{RS}. Such surface term also 
makes the variational principle to be well-defined and complete 
AdS space to be finite when brane goes to infinity.
We assume the surface terms in the following form \cite{NOch}:
\bea
\label{Iiv}
S_b&=&S_b^{(1)} + S_b^{(2)} \nn
S_b^{(1)} &=& \int d^d x \sqrt{\hat g}\left[
4\tilde a\hat R D_\mu n^\mu 
+ 2\tilde b_1 n_\mu n_\nu \hat R^{\mu\nu} D_\sigma n^\sigma 
+ 2\tilde b_2\hat R_{\mu\nu}D^\mu n^\nu \right. \nn
&& \left. 
+ {2 \over \tilde \kappa^2}D_\mu n^\mu \right] \nn
S_b^{(2)} &=& - \eta\int d^4 x \sqrt{\hat g} \ .
\eea
Here $n^\mu$ is 
the unit vector normal to the boundary.  
One can choose $\tilde b_1 = \tilde b_2$ 
but as we will see later, it is convienient to treat them as 
independent parameters when one considers the black hole background. 

We now choose the background as 
\be
\label{i}
ds^2=\e^{2\rho}dt^2 + \e^{-2\rho}dr^2 
+ r^2\sum_{i,j}^{d-1} g_{ij}dx^i dx^j\ ,
\ee
and we assume $g_{ij}$ is the metric of Einstein manifold, 
which is defined by $r_{ij}=kg_{ij}$, where $r_{ij}$ is 
the Ricci tensor and $k$ is a constant. For example, 3 
dimensional sphere corresponds to positive $k$ and hyperboloid 
to negative $k$. 
Here $\rho$ is treated as a dynamical variable, which can be 
determined as $\rho=\rho_0$ in (\ref{SAdS}) by solving the equations 
of motion. 
Then the curvatures, normal vector $n^\mu$ and its covariant 
derivatives are given by
\bea
\label{cvtrRcc}
R_{tt}&=&-\left(\rho'' + 2\left(\rho'\right)^2 
+ {(d-1)\rho' \over r}\right)\e^{4\rho} \nn
R_{rr}&=&-\rho'' - 2\left(\rho'\right)^2 
- {(d-1)\rho' \over r} \nn
R_{ij}&=&\left\{\left(-2r\rho' - d + 2\right)\e^{2\rho} + k 
\right\}g_{ij} \nn
0&=&\mbox{other Ricci tensors} \\
\label{crvrsclr}
R&=&\left(-2\rho'' - 4\left(\rho'\right)^2 
- {4(d-1)\rho' \over r} - {(d-2)(d-1) \over r^2}
\right)\e^{2\rho} \nn
&& + {(d-1)k \over r^2} \\
\label{nrmlvctr}
n^r&=&\e^\rho\ ,\quad \mbox{other components}=0 \nn
&& D_rn^r=0\ ,\quad D_tn^t=\e^\rho \rho'\ , 
D_i n^j = {\e^\rho \over r}\delta^i_{\ j} \nn
&& D_\mu n^\mu =\e^\rho \rho' + {(d-1)\e^\rho \over r} \nn
&& \mbox{other covariant derivatives}=0 \ .
\eea
The variations of the action (\ref{vi}) on the boundary, 
which lies at $r=r_0$,  
and  of (\ref{Iiv}) have the following forms
\bea
\label{viB}
\lefteqn{\left.\delta S\right|_{r=r_0}
=\int d^4x r_0^{d-1}\e^{2\rho}} \nn
&&\times\left[ 2a \left\{ -2 \delta\rho' + \delta\rho
\left(-8\rho' - {4(d-1) \over r_0}\right)\right\} \right. \nn
&&\times \left\{\left(-2\rho'' - 4\left(\rho'\right)^2 
- {4(d-1)\rho' \over r_0} - {(d-2)(d-1) \over r_0^2}
\right)\e^{2\rho} + {(d-1)k \over r_0^2} \right\} \nn
&& + 4b \left\{ \left\{ \delta\rho' 
+ \left(4\rho' + {d-1 \over r_0}\right)\delta\rho\right\}
\left(\rho'' + 2\left(\rho'\right)^2 
+ {(d-1)\rho' \over r_0}\right)e^{2\rho } \right. \nn
&& \left.-4(d-1)r_0^{-3} \delta\rho
\left\{\left(-2r\rho' - d + 2\right)\e^{2\rho} + k 
\right\}\right\} \nn
&& \left. + {1 \over \kappa^2}\left\{ -2 \delta\rho' + \delta \rho
\left(-8\rho' - {4(d-1) \over r_0}\right)\right\} \right] \\
\label{IviB}
\delta S_b&=&\int d^4x r_0^{d-1}\e^{\rho} \nn
&\times&\left[ \e^{3\rho}\delta\rho''\left\{\left(-8\tilde a 
 -2\tilde b_1\right)\left(\rho' + {d-1 \over r_0}\right)
 -2\tilde b_2\rho' \right\} \right. \nn
&& + \delta\rho'\left\{\tilde a \left\{\left( -32\rho' 
 - {16(d-1) \over r_0}\right)\left(\rho' + {d-1 \over r_0}\right)
 \e^{3\rho} \right.\right. \nn
&& + 4\left( \left(-2\rho'' - 4\left(\rho'\right)^2 
- {4(d-1)\rho' \over r_0} - {(d-2)(d-1) \over r_0^2}
\right)\e^{2\rho} \right. \nn
&& \left.\left. + {(d-1)k \over r_0^2} \right)\e^\rho
\right\} \nn
&& + \tilde b_1\left\{ \left( -8\rho' 
 - {2(d-1) \over r_0}\right)\left(\rho' + {d-1 \over r_0}\right)
 \e^{3\rho} \right. \nn
&& \left. + 2 \left(-\rho'' - 2\left(\rho'\right)^2 
 - {(d-1)\rho' \over r_0} \right)\e^{3\rho} \right\} \nn
&& + \tilde b_2 \left\{ 2 \left(-\rho'' 
 - 2\left(\rho'\right)^2 - {(d-1)\rho' \over r_0} \right)\e^{3\rho} 
 \right. \nn
&& \left.\left. - 8\left(\rho'\right)^2\e^{3\rho} 
 - {2(d-1) \over r_0}\rho'\e^{3\rho}
 - {4(d-1) \over r_0^2}\e^{3\rho} \right\}
 + {2\e^\rho \over \tilde \kappa^2}\right\} \nn
&& + \delta\rho\left\{ 4\tilde a\left\{4\e^{3\rho}
\left(-2\rho'' - 4\left(\rho'\right)^2 
 - {4(d-1)\rho' \over r_0} \right.\right.\right. \nn
&& \left.\left.\left. - {(d-2)(d-1) \over r_0^2}
\right)\left(\rho' + {d-1 \over r_0}\right) 
+ {2\e^\rho (d-1) k \over r_0^2}
\left(\rho' + {d-1 \over r_0}\right) \right\} \right. \nn
&& + 8\tilde b_1 \left(-\rho'' - 2\left(\rho'\right)^2 
 - {(d-1)\rho' \over r_0} \right)
\left(\rho' + {d-1 \over r_0}\right)\e^{3\rho} \nn
&& + 2\tilde b_2\left\{4\rho'\left(-\rho'' - 2\left(\rho'\right)^2 
 - {(d-1)\rho' \over r_0} \right)\e^{3\rho} \right.\nn
&& \left. + {4(d-1) \over r_0^3}
\left(-2r\rho' - d +2\right)\e^{3\rho} \right. \nn
&& \left.\left.\left. + {2k(d-1) \over r^3}\right\} 
+ {4\e^\rho \over \tilde\kappa^2} 
\left(\rho' + {d-1 \over r_0}\right) - \eta \right\}\right]\ .
\eea
In order that the variational principle is well-defined, 
the coefficients of $\delta\rho''$ and $\delta\rho'$ must vanish. 
 From the coeficient of $\delta\rho''$, one gets
\be
\label{abb}
\tilde b_1= -4 \tilde a\ ,\quad \tilde b_2=0\ .
\ee
Then substituting the black hole solution 
(\ref{SAdS}) into (\ref{viB}) and (\ref{IviB}), one finds 
\bea
\label{vr2}
\lefteqn{\delta S + \delta S_b} \nn
&=& \int d^d x r_0^{d-1}\e^{2\rho} \left[ \delta\rho'
\left\{\left(
{8d(d+1)a \over l^2} + {4db \over l^2} - {2 \over \kappa^2} 
\right.\right.\right. \nn
&& \left.\left. - {4d(d-1)\tilde a \over l^2} 
+ {2 \over \tilde\kappa^2}\right) - 4(d-1)
\left({d\e^{2\rho} \over r_0^2} + {k \over r_0^2} 
+ {d \over L^2}\right)\tilde a\right\}
\nn
&& + \delta\rho\left\{2\left(-8\rho' - {4(d-1) \over r_0}\right)
\left(- {d(d+1) \over l^2}\right)a \right. \nn
&& + \left(4\left(4\rho' + {d-1 \over r_0}\right){d \over l^2}
+ {4d(d-1) \over l^2 r_0}\right)b 
+ {1 \over \kappa^2}\left(-8\rho' - {4(d-1) \over r_0}\right) \nn
&& + 8 (d-1)\left(-{k \over r_0^2} 
- {2d \over l^2}\right)\left(\rho' + {d-1 \over r_0}\right)\tilde a \nn
&& \left.\left. 
+ {4 \over \tilde\kappa^2}\left(\rho' + {d-1 \over r_0}\right)
-\eta \e^{-\rho} \right\}\right]\ .
\eea
Then from the condition that the coefficient of $\delta\rho'$ 
vanishes, one obtains
\be
\label{cndtn}
\tilde a=0\ ,\quad 
0={8d(d+1)a \over l^2} + {4db \over l^2} - {2 \over \kappa^2} 
 + {2 \over \tilde\kappa^2} \ ,
\ee
which tells all the higher derivative terms in (\ref{Iiv}) 
should vanish. 
On the other hand, from the coefficient of $\delta\rho$, one has the 
equation of the motion on the boundary (here we put $\tilde a=0$):
\bea
\label{eqmtn}
0&=&\left({16d(d+1)a \over l^2} + {16db \over l^2} 
 - {8 \over \kappa^2} + {4 \over \tilde\kappa^2}\right)\rho' \nn 
&& + \left( {8(d-1)d(d+1) a \over l^2} + {8d(d-1)b \over l^2}
 - {8(d-1) \over \kappa^2} 
 + {4(d-1) \over \tilde\kappa^2}\right){1 \over r_0} \nn
&& - \eta \e^{-\rho}\ .
\eea
The parameter $\eta$ (brane tension) which is usually free parameter 
in brane-world cosmology is not free anymore and can be determined by the
condition that 
the leading divergence should vanish when one substitutes
the classical solution (\ref{SAdS})
into the action (\ref{vi}) with $c=0$ and into (\ref{Iiv}) with 
$\tilde a=\tilde b_{1,2}=0$:
\bea
\label{clactions}
S&=&\int d^4x r_0^d {1 \over d} \left\{{d^2(d+1)^2 a \over l^4}
+ {d^2(d+1)b \over l^4} - {d(d+1) \over \kappa^2 l^2}
 -\Lambda \right\} + {\cal O}\left(r_0^{d-1}\right) \nn
S_b&=&\int d^4x r_0^d \left\{ + {2d \over l^2 \tilde\kappa^2} 
 - {\eta \over l}\right\} + {\cal O}\left(r_0^{d-1}\right) \ .
\eea
Here (\ref{abb}) is used. Then one gets
\be
\label{eta1}
{\eta \over l}={d(d+1)^2 a \over l^4}
+ {d(d+1)b \over l^4} - {d+1 \over \kappa^2 l^2}
 - {\Lambda \over d} + {2d \over l^2 \tilde\kappa^2} 
\ee
or  deleting $\Lambda$ by using (\ref{ll})
\be
\label{eta2}
{\eta \over l}={4d(d+1)a \over l^4} + {4db \over l^4}
- {2 \over \kappa^2 l^2} + {2d \over l^2\tilde\kappa^2}\ .
\ee
We now consider the case of $d=4$. 
Then deleting $\tilde\kappa$ and $\eta$ in (\ref{eqmtn}), 
 using (\ref{cndtn}) and (\ref{eta2}), we obtain
\bea
\label{bdeq2}
0&=&\left({60b \over l^2} - {4 \over \kappa^2} \right) \rho' 
 + \left(-{480a \over l^2} - {12 \over \kappa^2} \right) {1 \over r_0} \nn
&& - \left(-{560a \over l^4} - {32b \over l^4} 
+ {6 \over \kappa^2 l^2} \right)l\e^{-\rho} \ .
\eea
Before substituting the explicit solution for $\rho$ in 
(\ref{SAdS}), we rewrite $\e^{2\rho}$ in the following form:
\bea
\label{rho}
&& \e^{2\rho} = {\left(r^2 - r_h^2\right) 
\left(r^2 + r_b^2\right) \over r^2l^2} \nn
&& r_h^2 = {l^2 \over 2}\left( - {k \over 2} 
+ \sqrt{{k^2 \over 4} + {4\mu \over l^2}}\right)\ ,\quad 
r_b^2 = {l^2 \over 2}\left(  {k \over 2} 
+ \sqrt{{k^2 \over 4} + {4\mu \over l^2}}\right)\ .
\eea
Here $r_h$ corresponds to the radius of the horizon. 
Then one can rewrite (\ref{bdeq2}) as follows:
\bea
\label{bdeq3}
0&=&\left({60b \over l^2} - {4 \over \kappa^2} 
\right) \left(-{1 \over r_0} + {r_0 \over r_0^2 - r_h^2} 
+ {r_0 \over r_0^2 + r_b^2} \right) 
+ \left(-{480a \over l^2} - {12 \over \kappa^2} 
\right) {1 \over r_0} \nn
&& - \left(-{560a \over l^4} - {32b \over l^4}
+ {6 \over \kappa^2 l^2} \right)
{r_0l^2 \over \sqrt{\left(r_0^2 - r_h^2\right)\left(r_0^2 + 
r_b^2\right)}} \ .
\eea
This equation determines $r_0$, that is, where the brane lies. 
When $r_0$ is large, the r.h.s. in (\ref{bdeq3}) behaves as
\be
\label{rhs1}
{\rm r.h.s.}\sim {1 \over r^2}
\left( {80a \over l^2} + {92 b \over l^2} 
- {22 \over \kappa^2} \right)\ .
\ee
and near the horizon ($r_0\sim r_h$), 
the r.h.s. in (\ref{bdeq3}) behaves as
\be
\label{rhs2}
{\rm r.h.s.}\sim {r_0 \over r_0^2 - r_h^2} 
\left({60b \over l^2} - {4 \over \kappa^2} \right)\ .
\ee
Therefore if 
\be
\label{cndtnr1}
\left( {80a \over l^2} + {92 b \over l^2} 
- {22 \over \kappa^2} \right)
\left({60b \over l^2} - {4 \over \kappa^2} \right)<0
\ , 
\ee
there is always a solution $r_0$, which satisfies (\ref{bdeq3}). 
Even if Eq.(\ref{cndtnr1}) is not satisfied, there can 
be solutions in general. 
For example, if $a = -{2b \over 25}$ and ${1 \over \kappa^2}
=-{16b \over 5l^2}$, we obtain from (\ref{bdeq3})
\be
\label{bdeq3b}
0={60b \over l^2} \left(-{1 \over r_0} + {r_0 \over r_0^2 - r_h^2} 
+ {r_0 \over r_0^2 + r_b^2} \right) - {16b \over l^4}
{r_0l^2 \over \sqrt{\left(r_0^2 - r_h^2\right)\left(r_0^2 + 
r_b^2\right)}} \ .
\ee
If we further consider the case where $k\rightarrow 0$ or 
$\mu\rightarrow + \infty$, we have 
$r_h^2\sim r_b^2 \sim l\mu^{1 \over 2}$. Then Eq.(\ref{bdeq3b}) 
reduces to
\be
\label{bdeq3bB}
15\left(r_0^4 + l^2 \mu \right) = 4 r_0^2 \sqrt{r_0^4 
+ l^2 \mu }\ ,
\ee
whose solutions are given by
\be
\label{bsol}
r_0^4= {225 \over 209} l^2\mu\quad \mbox{or}\quad l^2 \mu\ .
\ee
Since Eq.(\ref{bdeq3b}) contains $\sqrt{r_0^2 - r_h^2}$, $r_0$ 
should be larger than $r_h$. In the second solution in 
(\ref{bsol}), the brane approaches to the horizon. Nevertheless,
this indicates that brane observer lives outside of horizon of
multidimensional AdS BH. 

For the Einstein gravity case, where $a=b=\tilde a=0$, the 
condition (\ref{cndtnr1}) cannot be satisfied.  
Here, Eq.(\ref{cndtnr1}) has the following 
form:
\bea
\label{bdeq3E}
0&=& - {4 \over \kappa^2} 
\left({2 \over r_0} + {r_0 \over r_0^2 - r_h^2} 
+ {r_0 \over r_0^2 + r_b^2} \right) \nn
 && - {12 \over \kappa^2 r_0} 
+ {2 \over \kappa^2 l^2} 
{r_0l^2 \over \sqrt{\left(r^2 - r_h^2\right)\left(r_0^2 + 
r_b^2\right)}} \ .
\eea
Since r.h.s. in (\ref{bdeq3E}) is always negative, 
there is no solution $r_0$ which satisfies (\ref{bdeq3E}) 
for the Einstein gravity. However, modifications of approach
(for example, adding the arbitrary tension brane) may help
in construction of such d5 BH even in the Einstein gravity. 
Of course, in this case the radius of brane is not defined dynamically.

In general, one can get a solution without including the quantum 
effects from the matter fields on the brane \cite{sergio}\footnote{It could
be really interesting to estimate the role of such quantum
effects,especially in the study of Hawking radiation and related issues}. 
The brane has the topology of S$_1\times$M$_3$, where S$_1$ 
corresponds to the Wick-rotated time and M$_3$ is the 3 
dimensional Einstein manifold. Then, the proposal could be that for observer 
living on the brane the whole Universe evolution occurs within less than
one time period. Then, such "long-living" observer can realise 
that he lives outside of multidimensional BH horizon only by some clever
but indirect experiment. 

 If we Wick re-rotate the 
time coordinate, we obtain the brane universe, whose topology 
is given by R$_1\times$M$_3$, where R$_1$ corresponds to the 
time coordinate. Then the obtained universe is static. 
In the case $k>0$, however, one has another way of the Wick 
re-rotation. When $k>0$, one can choose M$_3$ as 3 dimensional 
unit sphere $S_3$, whose metric is given by 
\be
\label{S3mtrc}
ds_{{\rm S}_3}^2= d\theta^2 + \sin^2\theta d\Omega_2^2 \ .
\ee
Here $d\Omega_2^2$ corresponds to the metric of 2 dimensional unit 
sphere. If we change the variable by
\be
\label{chng}
\sin\theta = {1 \over \cosh\sigma}\ ,
\ee
the metric in (\ref{S3mtrc}) can be rewitten as
\be
\label{S3mtrcB}
ds_{{\rm S}_3}^2= {1 \over \cosh^2\sigma}\left(
d\sigma^2 + d\Omega_2^2\right) \ .
\ee
Then if one Wick re-rotate $\sigma$ by $\sigma\rightarrow 
-i\sigma$, S$_3$ becomes 3 dimensional de Sitter space. 
Then we obtain the brane universe, whose topology is 
S$_1 \times$(3d de Sitter space). It corresponds the  
inflationary universe, whose rate of the expansion is given by 
${1 \over r_0}$. Thus, we presented example of d5 S-AdS BH where 
4d observable Universe may be dynamically realized outside of horizon as
brane. Of course, the details of such scenario should be investigated but
as it looks we got preliminary positive answer to the question addressed in
the title.

It is interesting that 
the string theory dual to ${\cal N}=2$ superconformal field 
theory is presumably IIB string on ${\rm AdS}_5\times
X_5$ \cite{AFM} where $X_5=S^5/Z_2$. (The ${\cal N}=2$ $Sp(N)$ 
theory arises as the low-energy theory on the world volume on 
$N$ D3-branes sitting inside 8 D7-branes at an O7-brane). 
Then in the absence of Weyl term, 
${1 \over \kappa^2}$ and $\Lambda$ are given by 
\be
\label{prmtr}
{1 \over \kappa^2}={N^2 \over 4\pi^2}\ ,\quad
\Lambda= -{12N^2 \over 4\pi^2}\ .
\ee
This defines the bulk gravitational theory dual to super YM theory 
with two supersymmetries.
The Riemann curvature squared term in the above bulk action 
may be deduced from heterotic string via heterotic-type I duality 
\cite{Ts2}, which gives ${\cal O}(N)$ correction:
\be
\label{abc}
a=b=0\ ,\quad c= {6N \over 24\cdot 16\pi^2}\ .
\ee
Hence, HD gravity with above coefficients defines SG dual 
of super Yang-Mills theory (with two supersymmetries) in
next-to-leading order of AdS/CFT correspondence \cite{AdS}.
Note that in such version of d5 HD gravity S-AdS BH solution
may be constructed only perturbatively \cite{SNO}.Nevertheless,
it is very interesting to
 extend results of this work for such model with 
$c\neq 0$ because it has natural stringy basis as warped compactification
(with dynamical brane)
in next-to-leading order of AdS/CFT correspondence. This will be presented
elsewhere.

\ 

\noindent
{\bf Acknoweledgements.} The work by SDO has been supported in part by 
CONACyT(CP, ref.990356 and grant 28454E).


\end{document}